\documentclass[12pt]{article}

\begin{document}

\title{Spin 1/2 as propagation on a lattice with symmetries modulo 
       gauge transformations}
\author{L. Polley}
\date{\normalsize
FB Physik, Oldenburg University, 26111 Oldenburg, FRG}
\maketitle
\thispagestyle{empty}
\vspace*{-5mm}
\begin{abstract}
Relativistic spin 1/2, as represented by Susskind's 1977 discretization of 
the Dirac equation on a spatial lattice, is shown to follow from basic, not 
typically relativistic but essentially quantum theoretic assumptions: that 
position eigenstates propagate to nearest neighbours while respecting 
lattice symmetries modulo gauge transformations.
\end{abstract}
\setcounter{tocdepth}{2}
{\small\tableofcontents}
\newpage 
\setcounter{page}{1}

\section{Introduction}

The mutual consistency of quantum mechanics and special relativity 
has remained a nontrivial issue, particularly with respect to locality
\cite{Stapp} and quantum measurements \cite{Nikolic}, but also in 
describing a free particle. Quantum mechanics, on the one-particle level, makes 
a fundamental distinction between the roles of space 
and time which, it seems, can be overcome only in the framework of quantum
field theory. But even in quantum field theory, in the usual line of argument,
relativity is something that needs to be enforced. Moreover, if it {\em is}
enforced by postulating unitary representations of the Poincar\'e group,
there appear theoretical possibilities \cite{Shirokov} which are never observed 
in Nature, such as continuous spin\footnote{Continuous spin arises if,
in terms of induced representations, the little group of a light-like 
four-momentum is represented non-trivially in all of its components.} or 
tachyons.   

In this paper I suggest an ahistorical route to relativistic quantum 
mechanics as represented by the Dirac equation. I derive the equation,
and with it the Lorentz invariance, from seemingly ``non''-relativistic quantum 
theory. In fact, Dirac himself came to the conclusion in the 1950s \cite{Dirac51}
that the Michelson-Morley experiment, in view of symmetries present in 
quantum but not classical mechanics, had been overinterpreted as a support 
of special relativity. As for a derivation of Lorentz invariance from a 
``mechanism'', there is a well-known precedent: 
Maxwell \cite{Maxwell}, in a
balance of working hypothesis and actual belief \cite{Goldman}, utilized 
mechanistic ideas of electromagnetic fields which did not enforce relativity 
but got it right automatically. More recently, in constructing cellular 
automata, Bialynicki-Birula \cite{Bialynicki} noted that an automaton simulating
the Weyl equation would require only very general conditions: a two-component 
wave function, an evolution that is linear and unitary, and (a vague remnant of 
relativity) that a wave function constant in space be also constant in time. 

By the technicalities used (not by the direction of argument) the present 
paper is based on a discretization of the Dirac equation devised by Susskind 
in 1977 \cite{Susskind}. The issue was to compensate for the doubling of the 
degrees of freedom encountered in replacing a derivative with an 
antihermitian difference
$$
    \frac{\Delta f}{\Delta x} = \frac{f(x+a)-f(x-a)}{2a}
$$ 
A zero difference function, for example, is not only obtained from 
$f={\rm const}$ but also from an alternating constant $(-1)^{x/a}$ on the 
lattice sites. Susskind showed that spinorial degrees of freedom can be  
consistently assigned to different sites on a 3-dimensional lattice, thus 
thinning 
out the degeneracy of energy levels on a given lattice by a factor of four.
The discretized Dirac equation resulting in this way is, in case of zero mass,  
\begin{equation} \begin{array}{rcl} \label{hopDir}
i \dot{\psi}(x,y,z,t) & = & 
      i \left( \psi(x+1,y,z,t) -  \psi(x-1,y,z,t)\right) \\ 
      & + & i \left( \psi(x,y+1,z,t) -  \psi(x,y-1,z,t) \right) (-1)^x \\
      & + & i \left( \psi(x,y,z+1,t) -  \psi(x,y,z-1,t) \right) (-1)^{x+y} 
\end{array}  
\end{equation}
where $\psi(x,y,z,t)$ is a {\em one-component} wave function. Thus spin 1/2, 
usually thought of as ``internal'' to a point particle, can be encoded in a 
spatial arrangement of hopping amplitudes for a particle  without internal 
structure.   

In fact, equation (\ref{hopDir}) is the unique consequence of basic, not 
typically relativistic assumptions on the propagation 
of quantum particles living on the sites (as opposed to links 
or plaquettes) of a cubic spatial lattice:
\begin{itemize}
\item Locality: immediate propagation to nearest neighbours only
\item Lattice symmetries are realised {\em modulo gauge transformations}
\end{itemize}
Also, time evolution will be assumed to be linear and unitary. There will
remain two kinematical options, one of which will be discarded because it is 
infinitely slower than the other. Remarkably, the slow option is the one that 
would realise lattice symmetries in a strict sense, without accompanying gauge 
transformations.  

In Section 2 the assumptions are specified; they include a general 
equation for linear, unitary propagation as it was already proposed by this 
author \cite{Polley}. In Section 3, the 
assumed invariances of the equation of propagation are evaluated, and equation 
(\ref{hopDir}) is derived. As for introducing particle mass, it is pointed out 
that an alternative to the standard term due to Susskind exists which avoids 
species doubling on infinite lattices. In Section 4, I present my Conclusions.
In the Appendix, some omissions and simplifications anticipated in Sections 2 
and 3 are justified.  

\section{Specifying the assumptions}

\subsection{Nearest-neighbour hopping}

We assume, as in \cite{Polley}, that a quantum particle, initially in a position
eigenstate, will ``move'' by gradually (differentiably in $t$) forming 
superpositions of nearest-neighbour eigenstates. Then a general state vector, 
given as a superposition 
of eigenstates with coefficients $\psi(\vec{s},t)$, will evolve according to
\begin{equation} \label{hopEq}
    i \dot{\psi}(\vec{s},t) = \sum_{\vec{n}} \kappa(\vec{s},\vec{n}) \, 
           \psi(\vec{s}+\vec{n},t)
\end{equation}
where the sum runs over nearest neighbours, and where  
$\kappa(\vec{s},\vec{n})$ are complex hopping amplitudes whose properties are
to be determined.  The sum also includes an on-site hopping amplitude 
represented by $\vec{n}=0$.

Since the distances between nearest neighbours are all the same,
we assume that all hopping amplitudes are of the same magnitude. With
a suitable rescaling of time we thus assume
\begin{equation} \label{|kappa|}
      | \kappa(\vec{s},\vec{n}) | = 1 \quad\mbox{for all }\vec{s}
                                           \mbox{ and all }\vec{n}\neq 0
\end{equation} 
Unitarity of time evolution and hermiticity of the Hamiltonian
will be taken for granted. Using the standard scalar product of wave functions,
$$
   \langle \psi | \varphi \rangle = \sum_{\vec{s}} 
     \overline{\psi(\vec{s})} \ \varphi(\vec{s})
$$
the linear operator acting on the {\sc rhs} of (\ref{hopEq}) is hermitian if 
and only if
\begin{equation} \label{unitarity}
    \kappa(\vec{s},-\vec{n}) = \overline{\kappa(\vec{s}-\vec{n},\vec{n})}
\end{equation}

\subsection{Invariances modulo gauge transformations}

For a free particle, the equation of motion should be ``the same'' at all 
times and locations, as well as after a rotation. Quantum mechanically, the 
arbitrariness of the phases of position eigenstates allows to interpret 
``the same'' as ``gauge equivalent''.

In a local gauge transformation, the wave function at each space-time point 
is multiplied by a phase factor. Thus
$$
    \psi(\vec{s},t)_{\rm old} = g(\vec{s},t) \, \psi(\vec{s},t)_{\rm new} 
    \qquad |g(\vec{s},t)| = 1
$$
In terms of the new wave function, equation (\ref{hopEq}) involves the hopping 
amplitudes 
\begin{eqnarray}  \label{gaugeKappa}
    \kappa(\vec{s},\vec{n})_{\rm new} & = &  g(\vec{s}+\vec{n},t) \,
    \kappa(\vec{s},\vec{n})_{\rm old} \, g(\vec{s},t)^{-1}
    \qquad \vec{n}\neq 0 \\
    \kappa(\vec{s},0)_{\rm new} & = & \kappa(\vec{s},0)_{\rm old} 
              -i \dot{g}(\vec{s},t) g(\vec{s},t)^{-1} \label{gaugeKappa0}
\end{eqnarray}
If $S$ is a symmetry operation on the lattice (translation or rotation)
the hopping amplitudes of equation (\ref{hopEq}) would in general change
according to 
$$
  \kappa(\vec{s},\vec{n})_{\rm new} = \kappa(S^{-1}\vec{s},S^{-1}\vec{n})_{\rm old}
$$
Our assumption is that $\kappa_{\rm new}$ is a local gauge transform 
of $\kappa_{\rm old}$:
$$
   \kappa(\vec{s},\vec{n})_{\rm new} = g(\vec{s}+\vec{n}) ~ 
   \kappa(\vec{s},\vec{n})_{\rm old} ~ g(\vec{s})^{-1}
$$
Expressing this entirely in terms of $\kappa_{\rm old}$, and dropping the index, 
we have 
\begin{equation}   \label{gaugeS}
 g(\vec{s}+\vec{n})~\kappa(\vec{s},\vec{n})~g(\vec{s})^{-1}
                   =\kappa(S^{-1}\vec{s},S^{-1}\vec{n}) 
\end{equation}
Since $g$ depends on $S$ we eventually write $g(\vec{s},S)$.

\subsection{Maximal gauge fixing\label{secGauFix}}

Working in a particular gauge will greatly faciliate the evaluation of 
symmetries up to gauge transformations. Following the procedure of maximal
gauge fixing as devised in \cite{Duncan} for Hamiltonian lattice gauge theories,
let us choose to have 
\begin{equation} \label{kappa1fix}
   \kappa(x,y,z,\hat{1}) = 1 \mbox{ for all } x,y,z
\end{equation} 
This is accomplished, using (\ref{|kappa|}),  by a gauge 
transformation with a suitable behaviour in the $\hat{1}$ direction: 
$$
    g(x+1,y,z,t) =  g(x,y,z,t) \, \kappa(x,y,z,\hat{1})_{\rm old}^{-1}
$$
The values of $g$ on a plane with a constant $x$ coordinate are still free,
and can be used to fix $\kappa(\vec{s},\hat{2})$ on that plane. Let us choose to
have 
\begin{equation} \label{kappa2fix}
   \kappa(0,y,z,\hat{2}) = 1 \mbox{ for all }y,z
\end{equation}
which requires
$$
    g(0,y+1,z,t) =  g(0,y,z,t) \, \kappa(0,y,z,\hat{2})_{\rm old}^{-1}
$$ 
Finally, the values of $g$ along the line $x=y=0$ can be chosen so that 
\begin{equation} \label{kappa3fix}
    \kappa(0,0,z,\hat{3})  = 1 
\end{equation}
Any further gauge transformation $g$ that is not constant throughout the lattice
will destroy at least one of the conditions (\ref{kappa1fix})-(\ref{kappa3fix}).

\section{Propagation on a simple cubic lattice}

\subsection{Symmetries used}

For a translation by a vector $\vec{a}$ we have
\begin{equation}    \label{translations}
 S^{-1} \vec{n} = \vec{n} \qquad \qquad S^{-1}\vec{s} = \vec{s}-\vec{a}
\end{equation}
For a rotation by $90^{\circ}$ about the $\hat{1}$ axis, 
\begin{equation}    \label{xrotation}
\begin{array}{rcr}
 S^{-1} ~ \hat{1} & = &  \hat{1} \\
 S^{-1} ~ \hat{2} & = & -\hat{3} \\
 S^{-1} ~ \hat{3} & = &  \hat{2}
\end{array}        \qquad \qquad        S^{-1}(x,y,z) = (x,z,-y)
\end{equation}
For a rotation by $90^{\circ}$ about the $\hat{3}$ axis, 
\begin{equation}    \label{zrotation}
\begin{array}{rcr}
 S^{-1} ~ \hat{1} & = & -\hat{2} \\
 S^{-1} ~ \hat{2} & = &  \hat{1} \\
 S^{-1} ~ \hat{3} & = &  \hat{3}
\end{array}        \qquad \qquad        S^{-1}(x,y,z) = (y,-x,z)
\end{equation} 

\subsection{Determining the hopping amplitudes}

\subsubsection{Evaluating translations\label{Translations}}

Let $S$ in equation (\ref{gaugeS}) be a translation as specified in 
(\ref{translations}). Thus 
\begin{equation} \label{gaugeTau}
        g(\vec{s}+\vec{n},\vec{a}) \,
      \kappa(\vec{s},\vec{n}) \, g(\vec{s},\vec{a})^{-1}  =  
                                                 \kappa(\vec{s}-\vec{a},\vec{n})    
    \qquad \vec{n}\neq 0 
\end{equation}
Putting $\vec{n}=\hat{1}$ in (\ref{gaugeTau}) and using (\ref{kappa1fix}) we 
see that $g(\vec{s},\vec{a})$ must be independent of the coordinate $x$,
$$
   g(x,y,z,\vec{a}) = g(y,z,\vec{a})   \qquad\mbox{for all }\vec{a}
$$  
Now putting $\vec{n}=\hat{2}$ and $\vec{a}=\hat{1}$ in 
(\ref{gaugeTau}), we find
$$
 \kappa(x-1,y,z,\hat{2}) = \kappa(x,y,z,\hat{2})~
\left(g(y+1,z,\hat{1}) \, g(y,z,\hat{1})^{-1}\right)        
$$
Solving the recursion in $x$ and using gauge condition (\ref{kappa2fix}), we
have
$$
  \kappa(x,y,z,\hat{2}) = e^{ix\alpha(y,z)}
  \mbox{ where } e^{i\alpha(y,z)} = g(y,z,\hat{1}) \,
                                     g(y+1,z,\hat{1})^{-1}
$$ 
In fact, $\alpha$ must be independent of $y$ and $z$ since by re-inserting 
the last equation in (\ref{gaugeTau}) (with $\vec{n}=\hat{2}$) and considering 
$\vec{a}=\hat{2},\hat{3}$
we encounter an $x$ dependence of $\exp(ix\alpha(y,z))$ on the {\sc lhs}
and $\exp(ix\alpha(y-1,z))$ or $\exp(ix\alpha(y,z-1))$, respectively, on the 
{\sc rhs}. Hence, $\alpha(y-1,z)=\alpha(y,z-1)=\alpha(y,z)$ so that 
\begin{equation} \label{kappa2phi}
     \kappa(x,y,z,\hat{2}) = e^{i\alpha x} 
\end{equation} 
Using (\ref{kappa2phi}), equation (\ref{gaugeTau}) with $\vec{n}=\hat{2}$, 
$\vec{a}=\hat{1},\hat{2}$ can now be read as a recursion 
relation determining the gauge transformations $g(\vec{s},\hat{1})$ and 
$g(\vec{s},\hat{2})$ up to their values at $x=y=0$. We find
\begin{eqnarray} 
  g(y,z,\hat{1}) & = &  g(0,z,\hat{1}) \, e^{-i\alpha y} \label{trans1} \\
  g(y,z,\hat{2}) & = &  g(0,z,\hat{2})                   \label{trans2}
\end{eqnarray} 
To obtain restrictions on the hopping amplitude in the $\hat{3}$ direction, 
we now insert (\ref{trans1}) and (\ref{trans2}) into (\ref{gaugeTau}), using
$\vec{n}=\hat{3}$ and $\vec{a}=\hat{1},\hat{2}$. Thus
\begin{eqnarray*}   
 g(0,z+1,\hat{1})~\kappa(x,y,z,\hat{3})~g(0,z,\hat{1})^{-1}
                                &=&\kappa(x-1,y,z,\hat{3}) \\
 g(0,z+1,\hat{2})~\kappa(x,y,z,\hat{3})~g(0,z,\hat{2})^{-1}  
                               &=& \kappa(x,y-1,z,\hat{3})
\end{eqnarray*} 
Taking into account the gauge condition (\ref{kappa3fix}) the recursions are 
readily resolved, yielding
\begin{equation}
   \kappa(x,y,z,\hat{3}) =  e^{ix\beta(z)}~e^{iy\gamma(z)} \label{kappa3phi}
\end{equation}
where
\begin{eqnarray*}
    e^{i\beta(z)} & = &  g(0,z+1,\hat{1})^{-1}~g(0,z,\hat{1}) \\
   e^{i\gamma(z)}& = &  g(0,z+1,\hat{2})^{-1}~g(0,z,\hat{2})
\end{eqnarray*}

\subsubsection{Evaluating unitarity}

Applying (\ref{unitarity}) to (\ref{kappa1fix}), (\ref{kappa2phi}),
(\ref{kappa3phi}) we obtain
\begin{eqnarray}
     \kappa(x,y,z,-\hat{1}) & = & 1                                    \nonumber \\
     \kappa(x,y,z,-\hat{2}) & = & e^{-i\alpha x}               \label{kappa-phi} \\
     \kappa(x,y,z,-\hat{3}) & = & e^{-ix\beta(z-1)}~e^{-iy\gamma(z-1)} \nonumber
\end{eqnarray}

\subsubsection{Evaluating \boldmath$90^{\circ}$ rotations about the \boldmath$x$
               axis\label{xRotation}}

Let $S$ in equation (\ref{gaugeS}) be the rotation specified by (\ref{xrotation}). 
Putting $\vec{n}=\hat{1},\hat{2},\hat{3}$,
\begin{eqnarray}  
  g(x+1,y,z) ~ \kappa(x,y,z,\hat{1})~ g(x,y,z)^{-1} & = & \kappa(x,z,-y,\hat{1})
  \nonumber  \\
  g(x,y+1,z) ~ \kappa(x,y,z,\hat{2})~ g(x,y,z)^{-1} & = & \kappa(x,z,-y,-\hat{3})
  \label{gaugeRx} \\
  g(x,y,z+1) ~ \kappa(x,y,z,\hat{3})~ g(x,y,z)^{-1} & = & \kappa(x,z,-y,\hat{2})
  \nonumber
\end{eqnarray}
whence, using (\ref{kappa1fix}),(\ref{kappa2phi}), (\ref{kappa3phi}) and 
(\ref{kappa-phi}),
\begin{eqnarray*}
  g(x+1,y,z) ~ g(x,y,z)^{-1} & = & 1 \\
  g(x,y+1,z) ~ e^{i\alpha x} ~ g(x,y,z)^{-1} & = & 
                              e^{-ix\beta(-y-1)}~e^{-iz\gamma(-y-1)} \\
  g(x,y,z+1) ~ e^{ix\beta(z)}~e^{iy\gamma(z)}~ g(x,y,z)^{-1} & = & e^{i\alpha x}
\end{eqnarray*}
By the first of these equations, $g$ must not depend on $x$. Thus, in the third 
and second equation, the only $x$ dependence occurs in the exponentials, implying 
$$
    e^{i\beta(z)} = e^{i\alpha} = e^{-i\beta(-y-1)}\qquad \mbox{for all }y,z 
$$ 
Thus $\beta={\rm const}$, and there remain two possibilities,
\begin{equation}  \label{alpha=pi}
      e^{i\alpha}= e^{i\beta} = \pm 1 
\end{equation}

\subsubsection{Evaluating \boldmath$90^{\circ}$ rotations about the \boldmath$z$
               axis\label{zRotation}}

Now let $S$ in equation (\ref{gaugeS}) be the rotation specified by 
(\ref{zrotation}). Putting $\vec{n}=\hat{1},\hat{2}$,
\begin{eqnarray*}
  g(x+1,y,z) ~ \kappa(x,y,z,\hat{1})~ g(x,y,z)^{-1} & = & \kappa(y,-x,z,-\hat{2}) \\
  g(x,y+1,z) ~ \kappa(x,y,z,\hat{2})~ g(x,y,z)^{-1} & = & \kappa(y,-x,z, \hat{1})
\end{eqnarray*}
Using (\ref{kappa1fix}),(\ref{kappa2phi}),(\ref{kappa-phi}) we obtain
\begin{eqnarray*}
  g(x+1,y,z) ~ g(x,y,z)^{-1} & = &  e^{-i\alpha y} \\
  g(x,y+1,z) ~ g(x,y,z)^{-1} & = &  e^{-i\alpha x} 
\end{eqnarray*}
The solution to these recursion relations is 
$$
    g(x,y,z) = e^{-i\alpha xy} ~ g(0,0,z)
$$
The $xy$ dependent factor drops out when inserted in (\ref{gaugeS}) with 
$\vec{n}=\hat{3}$, leaving
$$
  g(0,0,z+1) ~ \kappa(x,y,z,\hat{3})~ g(0,0,z)^{-1} = \kappa(y,-x,z,\hat{3})
$$
Using (\ref{kappa3phi}) we arrive at
$$
   g(0,0,z+1)~e^{i\alpha x}~e^{iy\gamma(z)}~g(0,0,z)^{-1} 
                                                = e^{i\alpha y}~e^{-ix\gamma(z)}
$$
Considering the $x$ and $y$ dependences we obtain
\begin{equation}  \label{gamma=pi}
                e^{i\gamma} = e^{i\alpha} 
\end{equation}
Thus, using (\ref{alpha=pi}), the hopping amplitudes are determined up to the 
choice of
$ \alpha=0$,
\begin{equation}  \label{kappaScalar}
      \kappa(x,y,z,\hat{n}) \equiv 1
\end{equation}
or $\alpha=\pi$,
\begin{equation}  \label{kappaSpin} 
   \begin{array}{rcl}
      \kappa(x,y,z,\hat{1})~= & \kappa(x,y,z,-\hat{1}) & = ~ 1 \\
      \kappa(x,y,z,\hat{2})~= & \kappa(x,y,z,-\hat{2}) & =~ (-1)^x \\
      \kappa(x,y,z,\hat{3})~= & \kappa(x,y,z,-\hat{3}) & =~ (-1)^{x+y}\end{array}
\end{equation}

\subsection{Staticity of the scalar solution\label{staticity}}

Option (\ref{kappaScalar}) for the hopping amplitudes 
would also result from postulating {\em strict} invariance under the lattice
symmetries, as already studied in \cite{Polley}. Its continuum
limit was found to be the nonrelativistic Schr\"odinger equation of a scalar 
particle. However, the
time scale\footnote{$\kappa$ is 
an inverse time by equation (\ref{hopEq}). The lattice spacing $a$ emerges 
from Taylor expansions of next-neighbour terms; hence, its dimension is always 
cancelled by that of a spatial derivative. The derivatives act on wave functions
in the continuum limit, so they are independent of $a$.}
on which the wave functions would evolve was found to be 
$\lambda^2/\kappa a^2$, 
where $\kappa$ denotes the nearest-neighbour hopping amplitude, and $\lambda$
is a length scale of the wave function.
$\lambda^2/\kappa a^2$ is also the time scale of unitary cellular automata
simulating scalar particles \cite{Boghosian}. In contrast, the time scale
of option (\ref{kappaSpin}) can be seen to be $\lambda/\kappa a$ 
from the initial choice of scale in (\ref{|kappa|}) and its modification in 
(\ref{masslessDirac}); again, $\kappa$ denotes the nearest-neighbour hopping 
amplitude. If $a$ is very small (like the Planck length), we obviously have
$$
         (\lambda/a)^2 \gg \lambda/a
$$
so option (\ref{kappaScalar}) tends to a static 
(non-kinetic and, in this sense, non-particle) scenario relative to 
(\ref{kappaSpin}). 

\subsection{Recovering the massless Dirac equation}

This section reviews standard procedure with lattice fermions.
Using (\ref{kappaSpin}) the hopping equation (\ref{hopEq}) reads 
$$ \begin{array}{rcl}
i \dot{\psi}(x,y,z,t) & = & 
       \left( \psi(x+1,y,z,t) +  \psi(x-1,y,z,t)\right) \\ 
      & + & \left( \psi(x,y+1,z,t) +  \psi(x,y-1,z,t) \right) (-1)^x \\
      & + & \left( \psi(x,y,z+1,t) +  \psi(x,y,z-1,t) \right) (-1)^{x+y} 
\end{array}  
$$
Equation (\ref{hopDir}) is recovered by the gauge transformation 
$\psi_{\rm old} = i^{x+y+z}\psi_{\rm new}$. For the new $\psi$,
$$ \begin{array}{rcl} 
i \dot{\psi}(x,y,z,t) & = & 
      i \left( \psi(x+1,y,z,t) -  \psi(x-1,y,z,t)\right) \\ 
      & + & i \left( \psi(x,y+1,z,t) -  \psi(x,y-1,z,t) \right) (-1)^x \\
      & + & i \left( \psi(x,y,z+1,t) -  \psi(x,y,z-1,t) \right) (-1)^{x+y} 
\end{array}  
$$
Since the alternating sign factors are strongly fluctuating when viewed
on a length scale much larger than the lattice spacing $a$, there can be no 
smooth solution to equation (\ref{hopDir}). However, the equation is 
solved by a superposition of wave functions of the form
\begin{equation} \label{4solutions}
 \begin{array}{ll}
  \psi_{00}(x,y,z,t) \quad & \qquad 
  \psi_{01}(x,y,z,t) (-1)^y \\
  \psi_{10}(x,y,z,t) (-1)^x \quad & \qquad 
  \psi_{11}(x,y,z,t) (-1)^{y+x} \end{array} 
\end{equation}
where $\psi_{AB}(x,y,z,t)$ is assumed to be smooth in the sense that it 
varies from a lattice site to the next in ${\cal O}(a)$ at most.
As suggested by the double index, the space of solutions is a tensor product.
Multiplication by $(-1)^x$, for example, interchanges the presence/absence
of that factor in the wave function, hence it is represented by the 
Pauli matrix $\sigma_1$ acting on the first index, and by 
$\sigma_1\otimes{\bf 1}$ acting on both indices.
Similarly, differentiating along the $x$ direction gives an extra minus sign 
depending on whether the factor $(-1)^x$ is present or absent; this corresponds
to the action of $\sigma_3\otimes{\bf 1}$. Thus the right-hand side of 
(\ref{hopDir}) combines matrix factors and spatial differences into
\begin{equation} \label{sigmaXsigma}
   \sigma_3\otimes{\bf 1}~i\Delta_x
 + \sigma_1\otimes\sigma_3~i\Delta_y
 + \sigma_1\otimes\sigma_1~i\Delta_z
\end{equation} 
The tensor products are readily seen to satisfy the algebraic
relations of the Dirac $\alpha$ matrices. The difference operations
asymptotically tend to $2a\partial/\partial x$,
$2a\partial/\partial y$, $2a\partial/\partial z$ in the continuum 
limit $a\to 0$. We may absorb the factor of $2a$ in a redefinition of 
the time parameter, thus recovering the massless  Dirac equation
\begin{equation} \label{masslessDirac}
    i \frac{\partial\psi}{\partial t'} = 
                           i \alpha_k \frac{\partial\psi}{\partial x_k} 
   \qquad \qquad t' = 2at
\end{equation}  

\subsection{Mass terms\label{MassTerms}}

In spatial continuum, massless Dirac particles have chiral symmetry. This
symmetry gets broken if a mass term is introduced. Therefore it is not
unsatisfactory to find that mass terms on a lattice may require the 
breaking of a lattice symmetry. Susskind's mass term is an on-site hopping 
amplitude $\mu (-1)^{x+y+z}$ so that equation (\ref{hopDir}) becomes
$$
   i \dot{\psi}(x,y,z,t) = \left( i \Delta_x  
      + i (-1)^x \Delta_y
      + i (-1)^{x+y} \Delta_z 
      +  \mu(-1)^{x+y+z}\right) \psi(x,y,z,t)
$$
The last term breaks the invariance (modulo gauge transformations) under 
translations  by one lattice unit, while invariance under translations by 
two units is preserved. 
Susskind's mass term requires a doubling of the dimension of the space of 
solutions, since the functions (\ref{4solutions}) need to be complemented by 
analogous functions with an extra factor of $(-1)^z$. This expands the tensor 
products (\ref{sigmaXsigma}) of the Hamiltonian to
$$
   \sigma_3\otimes{\bf 1}\otimes{\bf 1}~i\Delta_x
 + \sigma_1\otimes\sigma_3\otimes{\bf 1}~i\Delta_y
 + \sigma_1\otimes\sigma_1\otimes\sigma_3~i\Delta_z
 +  \mu\sigma_1\otimes\sigma_1\otimes\sigma_1
$$
The enlarged space of states is also recovered by acting on (\ref{4solutions}) 
with a symmetry of equation (\ref{hopDir}), Susskind's version of the parity 
operation
$$
   \psi(x,y,z,t) \longrightarrow  (-1)^{x+y+z} \psi(-x,-y,-z,t)
$$   
In Monte Carlo simulations, which can only use lattices with a finite number 
of sites, there is no natural distinction between smooth and strongly 
fluctuating wave functions. Thus the degeneracy related to the above parity 
operation can only be suppressed at the expense of some arbitrariness.       

On infinite lattices, however, an extra factor of $(-1)^z$ to the functions
(\ref{4solutions}) does make a difference. It may therefore be of interest to 
note that Susskind's term is not the only possibility of introducing mass.
For example, we may allow for a variation of the {\em magnitude} of the hopping 
amplitude in the $x$ direction so as to violate (\ref{|kappa|}) while keeping 
(\ref{unitarity}), 
$$
    \kappa(x,y,z, \hat{1}) =  i + i\mu \, (-1)^x \qquad \qquad
    \kappa(x,y,z,-\hat{1}) = -i + i\mu \, (-1)^x
$$
Physically this would correspond to an alternating variation of the lattice 
spacing. The additional term in hopping equation (\ref{hopDir}) is 
$$
   i\, \mu \, (-1)^x \, \left(\psi(x+1,y,z,t)+\psi(x-1,y,z,t)\right)
$$
Repeating the arguments that lead to (\ref{sigmaXsigma}) we obtain the
operator 
$$
     \sigma_3\otimes{\bf 1}~i\Delta_x
 + \sigma_1\otimes\sigma_3~i\Delta_y
 + \sigma_1\otimes\sigma_1~i\Delta_z
 + 2i \mu (\sigma_1\sigma_3)\otimes{\bf 1}
$$
Since $i\sigma_1\sigma_3=\sigma_2$ and since $\sigma_2\otimes{\bf 1}$
anticommutes with the first three tensor products we recover the Dirac 
matrices in the form
$$
    \sigma_3\otimes{\bf 1} = \alpha_1 \qquad
    \sigma_1\otimes\sigma_3 = \alpha_2 \qquad
    \sigma_1\otimes\sigma_1 = \alpha_3 \qquad
    \sigma_2\otimes{\bf 1} = \beta
$$

\section{Conclusions}

We have derived Susskind's discretization of the Dirac equation from
assumptions which apparently do not anticipate special relativity. While
time was assumed to run continuously,
spatial coordinates were confined to a lattice (reminiscent of a 
stack of particle detectors). Locality, too, was imposed in an unrelativistic 
sense, assuming that propagation from some position will, within a  
short interval of time, reach the nearest neighbours only. The intrinsically 
quantum-mechanical assumption was that the amplitudes of propagation will 
respect the symmetries of the lattice {\em to the extent they have to} 
 in quantum theory, namely up to phase shifts of position eigenstates. 

On the basis of these assumptions, a non-relativistic and a relativistic option 
appeared at the same stage in section \ref{staticity}. As it happened, this was
simultaneously the alternative between strict symmetry and symmetry modulo gauge 
transformations. It was a matter of kinematical speed, rather than principle, 
that the non-relativistic option was discarded.  

What insight do we gain by this route to the Dirac equation? 
I think it explains the preferred role of spin 1/2 in the Standard Model, 
since {\em no} internal structure of a particle (of a kind living on 
lattice sites) was assumed, and yet the Dirac equation resulted. 
In particular, in the case of zero mass where 
continuous spin is a possibility consistent with Poincar\'e invariance, 
it was just the massless version of the Dirac equation which emerged.  
More generally, I think the unity rather than mere consistency of 
special relativity and quantum theory---even in a  ``non''-relativistic 
formulation of the latter---is emerging here. 
Finally, gauge transformations turn out to be as fundamental to the propagation 
of free particles as to particle interactions.  
 
It would be interesting to determine the ``internal'' degrees of freedom 
on lattices with other than simple cubic structure, especially with some of
the infinitely many close-pack structures \cite{Mermin}.

\appendix

\section*{Appendix}

\addcontentsline{toc}{section}{~~~~Appendix}

\subsection*{Time-dependence ruled out\label{appA}}

To simplify the notation, hopping amplitudes were so far considered as 
functions of the spatial coordinates only. Here we show that in the gauge we 
were using there is, in fact, no other possibility consistent with 
time-translation invariance modulo gauge transformations.

Let $g(\vec{s},t)$ be the gauge transformation accomplishing the shift $S$ of 
the hopping amplitudes by a time $\delta t$. Equation (\ref{gaugeS}) then reads 
\begin{equation}  \label{gaugeDt}
 g(\vec{s}+\vec{n},t)~\kappa(\vec{s},\vec{n},t)~g(\vec{s},t)^{-1}
                   =\kappa(\vec{s},\vec{n},t-\delta t) 
\end{equation}
Putting $\vec{n}=\hat{1}$ and using our gauge condition (\ref{kappa1fix}),
which holds at all times, we obtain $g(x+1,y,z,t)=g(x,y,z,t)$ and hence
$$
    g(x,y,z,t) = g(y,z,t)
$$
Now putting $\vec{n}=\hat{2}$ and $x=0$ in (\ref{gaugeDt}) and using gauge 
condition (\ref{kappa2fix}) we moreover obtain
$$
    g(y,z,t) = g(z,t)
$$
Finally, putting $\vec{n}=\hat{3}$ and $x=y=0$ and using gauge condition
(\ref{kappa3fix}) we see that the gauge factor can only be a function of the 
time parameter:
$$
    g(z,t) = g(t)
$$ 
But then $g$ has no effect at all in equation (\ref{gaugeDt}), and the
nearest-neighbour hopping amplitudes must be strictly invariant under a time
shift: 
$$
    \kappa(\vec{s},\vec{n},t) = \kappa(\vec{s},\vec{n},t-\delta t)
$$
 
\subsection*{On-site hopping gauged away} 

On-site hopping amplitudes consistent with the spatial symmetries
(modulo gauge transformations) can always be gauged away, as we 
now show. Thus it was justified to omit them in the previous sections
(excluding section \ref{MassTerms} where translational
symmetry was partially broken).  

The analogue of equation (\ref{gaugeS}), using gauge transformation 
(\ref{gaugeKappa0}), would be
\begin{equation}   \label{kappa0(t)}
   \kappa(\vec{s},0,t) -i  \dot{g}(\vec{s},t) g(\vec{s},t)^{-1} 
 = \kappa(S^{-1}\vec{s},0,t)
\end{equation}
Let us reconsider the gauge factors which accomplished spatial translations 
of the nearest-neighbour amplitudes (section \ref{Translations}), taking into 
account the final expressions of the amplitudes as given by (\ref{kappaSpin}).
Equation (\ref{gaugeTau}) with $\vec{n}$ put equal to $\hat{1},\hat{2},\hat{3}$
can then be read as a recursion relation determining the $x,y,z$ dependence of 
$g(x,y,z,\vec{a})$ for a given shift vector $\vec{a}$. The arguments leading 
to those expressions did anticipate that $\kappa(x,y,z,\hat{n})$ would be 
independent of time, but this was justified in the previous section.

Those recursion relations implicit in (\ref{gaugeTau}) determine
$g(x,y,z,\vec{a})$, at each instant of time, up to a global phase factor.
That factor could depend on $t$, allowing for an expression of the form 
$$
   g(x,y,z,t,\vec{a}) = g_0(x,y,z,\vec{a}) \cdot h(t,\vec{a})
$$ 
Inserting this in (\ref{kappa0(t)}) the time-independent $g_0$ drops out,
leaving
$$
    \kappa(\vec{s},0,t) -i \dot{h}(t,\vec{a}) h(t,\vec{a})^{-1} 
                                     = \kappa(\vec{s}-\vec{a},0,t)
$$
In particular, putting $\vec{a}=\hat{1},\hat{2},\hat{3}$ and defining
$$
     \vec{c}=(c_1,c_2,c_3)     \qquad \qquad 
     c_n(t) =  i \dot{h}(t,\hat{n}) h(t,\hat{n})^{-1}                             
$$
we see that on-site hopping amplitudes can at most take the form
\begin{equation} \label{kappa0cs}
\kappa(\vec{s},0,t)=\kappa(\vec{0},0,t)+\vec{s}\cdot\vec{c}(t) 
\end{equation}
But $\vec{c}(t)$ must vanish due to rotational symmetry (modulo 
gauge transformations) by an argument similar to the above for translations. 
Reconsidering section \ref{xRotation} we find that equations (\ref{gaugeRx}) 
in conjunction with (\ref{kappaSpin}) determine the gauge factor up to a 
time-dependent global factor,
$$
   g(x,y,z,t,R_x) = g_0(x,y,z,R_x)\cdot h(t,R_x)
$$ 
In equation (\ref{kappa0(t)}), again, the time-independent $g_0$ drops out,
leaving 
$$
    \kappa(\vec{s},0,t) -i \dot{h}(t,R_x) h(t,R_x)^{-1} 
                                     = \kappa(R_x^{-1}\vec{s},0,t)
$$    
Taken at the origin $\vec{s}=0$ the equation implies that $\dot{h}$ must 
vanish. Hence, $\kappa(\vec{s},0,t)$ must be {\em strictly} invariant 
under the $90^{\circ}$ rotation about the $x$ axis, which implies that 
$\vec{c}(t)$ can at most have an $x$ component. This latter possibility 
can finally be ruled out by reconsidering the $90^{\circ}$ rotation about 
the $z$ axis as in section \ref{zRotation}. 

The remaining term of (\ref{kappa0cs}) can be removed by a gauge transformation 
only dependent on time, satisfying $\dot{g}(t) =-i\kappa(\vec{0},0,t) g(t)$.

\end{document}